# Atomically-Precise, Custom-Design Origami Graphene Nanostructures

Hui Chen[1†], Xian-Li Zhang [1†], Yu-Yang Zhang[1,2†], Dongfei Wang[1], De-Liang Bao[1,2], Yande Que[1], Wende Xiao [1], Shixuan Du[1 *], Min Ouyang[3], Sokrates T. Pantelides[2,1], Hong-Jun Gao[1*]

[1]Institute of Physics & University of Chinese Academy of Sciences, Chinese Academy of Sciences Beijing 100190, China

[2]Department of Physics and Astronomy and Department of Electrical Engineering and Computer Science, Vanderbilt University, Nashville, Tennessee 37235, United States

[3]Department of Physics, University of Maryland, College Park, Maryland 20742-4111, United States

*Correspondence to:   hjgao@iphy.ac.cn, sxdu@iphy.ac.cn

†These authors contributed equally to this work.

**The construction of atomically-precise carbon nanostructures holds promise for developing novel materials for scientific study and nanotechnology applications. Here we show that graphene origami is an efficient way to convert graphene into atomically-precise, complex, and novel nanostructures. By scanning-tunneling-microscope manipulation at low temperature, we repeatedly fold and unfold graphene nanoislands (GNIs) along arbitrarily chosen direction. A bilayer graphene stack featuring a tunable twist angle and a tubular edge connection between the layers are formed. Folding single-crystal GNIs creates tubular edges with specified chirality and one-dimensional electronic features similar to those of carbon nanotubes, while folding bicrystal GNIs creates well-defined intramolecular junctions. Both origami structural models and electronic band structures were computed to complement analysis of the**

**experimental results. The present atomically-precise graphene origami provides a platform for constructing novel carbon nanostructures with engineered quantum properties and ultimately quantum machines.**

**One Sentence Summary:** We use a STM tip to create custom-design, complex graphene origami nanostructures, ultimately envisioning functional quantum machines.

The discovery of fullerenes, nanotubes, and, more recently, the isolation of monolayer graphene sparked a revolution in the fabrication of an extraordinary variety of $sp^2$-bonded carbon allotropes (*1, 2*). Graphene itself, along with variants that include five- and/or seven-member rings, can be viewed as building blocks of $sp^2$-bonded allotropes (*3*) and of three-dimensional graphene-based nanostructures (GNSs) and devices that have been either fabricated or predicted theoretically for potential applications (*4-7*), even machines (*8*). Experimental realization of GNSs has been pursued by a variety of chemical, electrochemical, mechanical, radiation-assisted, and other approaches (*2*). These approaches, however, lack the ability to produce pure, structurally uniform GNSs, with bottom-up chemical synthesis viewed by some as the most promising route toward that goal (*9*). Meeting this challenge for GNSs with sub-10-nm features that can be used for quantum functionalities is even more demanding, as atomic-level precision is needed.

Origami, the ancient art of paper folding, has been widely used in diverse areas, from architecture to battery design and DNA nanofabrication (*10*). It has also inspired the fabrication or simulation of macroscale origami graphene structures and devices (*11-20*), even machines (*21*). Nanoscale graphene origami, however, where quantum phenomena are expected to be manifest, has been mainly the realm of theoretical investigations, predicting GNSs with unusual physical properties such as an ability to carry spin-polarized currents for spintronic applications(*22*), fold-induced gauge fields (*23*), large permanent electric dipoles (*24*), strong magnetophotoelectric effect (*25*), and topologically protected fold states (*26*). The electronic properties of folded GNSs have been predicted to depend sensitively on detailed atomic configurations (*24, 27*). In addition, graphene sheets containing particular defects (*28*) can in principle be used to create origami GNSs with unique functionalization.

The experimental demonstration of atomically precise, nanoscale graphene origami, however, has received limited attention. "Graphite origami" was first envisioned in 1995 by Ebbesen and Hiura (*29*), who observed accidental tearing and folding of graphite surface layers by an atomic-force-microscope (AFM) tip. In 1998, Roy *et al.* (*30, 31*) showed that a scanning-tunneling-microscope (STM) tip can in fact be used to induce folding of "graphitic sheets" at step edges of graphite, but without control of the folding direction. Similar results were reported in 2006 by Li *et al.* (*32*) using an AFM tip. Tearing and deformations were observed. In 2008, Schniepp *et al.* (*33*) achieved folding and unfolding of monolayer graphene, but concluded that the folds occur at pre-existing kink or fault lines. More recently, Kim *et al.* reported folding and unfolding of graphene within a V-cut made on the top sheet of graphite, but the folding angle was constrained by the "pinning" effect of the graphene edge and the operation was typically accompanied by tears or damage (*34*). It is clear that atomically precise and controllable graphene origami for the creation of custom-design GNSs with quantum features remains an open challenge.

In this paper, we report the atomically-precise folding and unfolding of graphene GNIs on a HOPG substrate without any tears or damage, including two-dimensional (2D) stacked bilayer graphene with precisely tunable twisting angle, one-dimensional (1D) folded tubular edge which is associated with an intramolecular junctions (IMJs). More specifically, by using tip of a STM, we repeatedly fold GNIs to achieve origami GNSs and unfold them into their original topography. By controlling folding directions, various stacked bilayer GNSs possessing arbitrary twisting angle up to 60° with an accuracy of 0.1° can be fabricated with the unique feature that the bilayer is connected by a tubular edge. The chirality and corresponding electronic properties of as-formed one-dimensional tubular nanostructures on the edges, which resemble carbon nanotubes, are precisely controlled. Furthermore, by folding bi-crystal GNIs with atomically-well-defined domain boundaries, analogs of carbon nanotube (CNT) IMJs have been created and their electronic properties have been probed by scanning tunneling spectroscopy (STS) and compared with density-functional-theory calculations. Model GNSs were optimized using classical force fields and were used to complement the analysis of experimental images. The present work provides a route for the fabrication of GNSs with engineered properties and construction of graphene-based quantum machines. Furthermore, the results reported in this paper set the stage for the discovery of new and unusual phenomena as the folded GNIs are composite structures comprising a CNT-like fold and a twisted bilayer graphene (TBLG). For example, it may be worth exploring the superconductivity of the TBLG part with a magic twist angle, attached to either a semiconducting or metallic tube or an IMJ.

## *Results and Discussion*:

Figures 1A and 1B show a schematic graphic and experimental demonstration of STM origami by sequentially folding and unfolding a single GNI along a pre-defined direction, respectively. Briefly, in order to fold a GNI, a STM tip is brought close to its edge by reducing the tunneling resistance in the STM junction, followed by moving across the GNI along a pre-determined direction (arrows in both 1A and 1B). During its motion, the tip literally lifts the GNI by the edge, drags the GNI along the tip's own track, stops, and places the moving portion of the GNI at the desired location. This process results in a folded GNS, in which part of the GNI is vertically stacked upon the remaining part to form a 2D bilayer graphene stack that is connected by a 1D tubular edge (Fig. 1C). A reversed process can be performed to unfold the new GNS with full recovery of the original GNI, also shown in Fig. 1B. Such folding and unfolding processes can be repeated multiple times with the same GNI along arbitrary directions without causing damage or structural defects. It is, therefore, possible to achieve various desirable GNSs without changing their local environment to facilitate systematic structure-property studies. A few unique features of our STM nano-origami stand out as compared with all existing relevant literatures (*29-34*) and open up exciting opportunities for further study that cannot be achieved otherwise: (1) The folding operation is spatially localized, i.e., the operation on one GNI has no effect on its neighbors (fig. S1); (2) The folding direction is arbitrary and atomically precise; (3) There is no size limitation on the GNIs, which makes it feasible to create folded GNSs at different length scales (fig. S2); and (4) No damage or structural defect has been induced during this repeatable process.

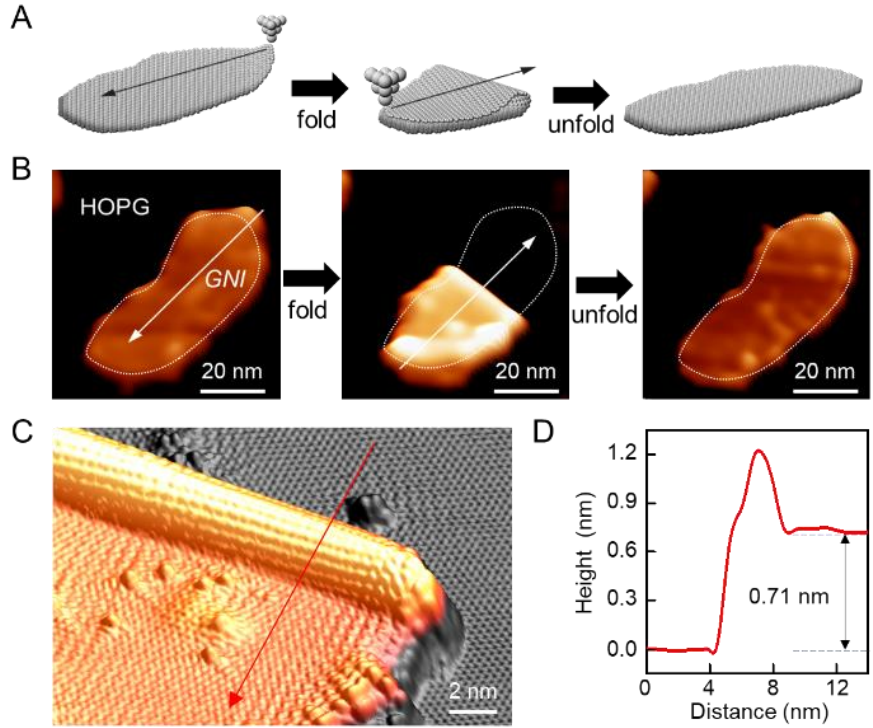

**Fig.1. Construction of atomically well-defined folded GNSs by STM origami.** (**A**) Schematic graphic of folding and unfolding a GNI along an arbitrary direction defined by the black arrow. (**B**) Experimental realization of (A). Series of STM images showing a sequence of the folding and unfolding of a GNI along the direction indicated by the white arrows. (**C**) Three-dimensional STM topography of a typical folded GNS. (**D**) Line profile along the red arrow in (C), showing formation of both the 1D tubular edge and the 2D stacked graphene flatland with height comparable to the distance between two graphene layers (0.70 nm). Settings for (B): tunneling current $I_t$ = 10 pA, bias voltage $V_s$= −3 V; Setting of (C) $I_t$ =100 pA, $V_s$ =1 V. The GNIs were manipulated by using lateral tip-induced manipulation with a typical current of ~ 100 pA and a voltage of ~ 3 mV. All results were acquired at T = 4.2 K.

The STM-origami GNSs are of high quality and atomically well-defined. Figure 1C shows atomic arrangements of one typical folded GNS, highlighting two different structural features: a 2D bilayer flatland and a 1D tubular structure on the edge. A line profile across the folded GNS is shown in Fig. 1D. The bilayer nature of the GNS is confirmed by the height between the top layer and the substrate, namely ~ 0.71 nm, which is comparable to the distance between two graphene layers (~ 0.70 nm) (*35*). The curved profile of the edge, which is higher than the flat top layer, corroborates its identification as a tubular edge.

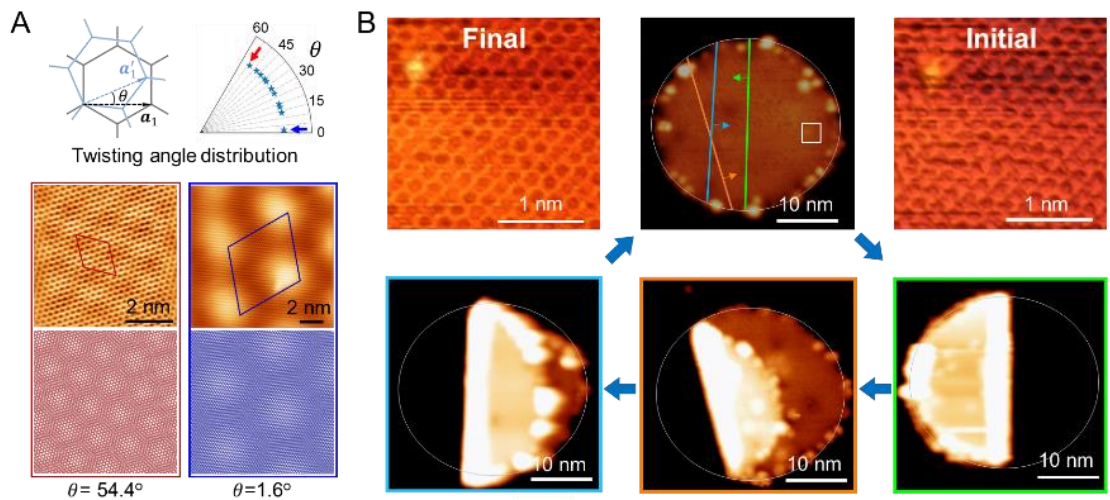

**Fig. 2. Precisely controlled folding of a GNI along preselected directions.** (**A**) Top-left: Schematic graphic of a 2D bilayer GNS with twisting angle $\theta$ produced by folding a GNI. The angle $\theta$ can be uniquely defined by the folding direction depicted in Figures 1A and 1B (see also fig. S3). Light-blue (lattice vector $a'_1$) and black (lattice vector $a_1$) hexagonal lattices represent top and bottom layers, respectively. Top-right: Summary of all experimental $\theta$ values achieved in the current work, showing the tunable range and precise control of $\theta$. Bottom: STM images and corresponding models showing moiré patterns of two exemplary folded GNSs with different $\theta$, 54.4° and 1.6°, respectively. These two

$\theta$ value correspond to the red and blue arrow highlights in the top-right, respectively. Periodic cells are marked by red and blue rhombuses in the left and right panels, respectively. (**B**) Series of STM images showing repeatable folding and unfolding of a single GNI along different directions by STM origami. Three examples of folded GNSs from the same GNI (top-middle) but different folding directions are shown in the bottom panels. The color frame of bottom images corresponds to their folding axis with the same color code marked in the original GNI in the top-middle panel. The atomically resolved images in the top-right and top-left panels were acquired from the white square area in the top-middle image before and after multiple origami operations, respectively, showing no occurrence of structural damage. Settings for (A): $I_t$ = 100 pA, $V_s$ = -0.1 V; Settings for (B): $I_t$ = 100 pA, $V_s$ = -0.1 V (top-right and top-left); $I_t$ = 10 pA, $V_s$ = -3.0 V (others).

We have repeated the folding and unfolding of a GNI along several directions sequentially. The arbitrary folding capability demonstrated in Fig. 1 immediately opens up an exciting opportunity to achieve graphene stacking with a tunable twist (Fig. 2A). The twisting angle $\theta$ between the top layer and the bottom layer can be uniquely determined by the folding direction (i.e. the moving direction of the STM tip for the origami operation) (fig. S3). Figure 2B (bottom) shows three exemplary GNSs with distinct folding orientations from the same GNI (top-middle). It is worth noting that in addition to bilayer stacking, by sequentially folding for multiple times, multilayer stacked GNSs can also be obtained (fig. S4). Importantly, we have found that, even after multiple folding and unfolding steps, the overall morphology of GNI remains the same without appearance of defects, as determined by comparing high-resolution STM images recorded before and after operations (Fig. 2B). These data suggest that STM origami is a safe and gentle operation that is essential for the construction of high-quality stacked structures.

Direct evidence of stacking with different twist angles is the formation of a tunable moiré superstructure on folded GNS. Figure 2A (bottom) presents typical high-resolution STM images and corresponding models of two different GNSs formed by folding the same GNI along different directions. The folding angles (fig. S3) utilized for creating these two GNSs are 0.8° and 27.2°, lead to the resultant $\theta$ of 1.6° and 54.4°, respectively. This estimation of $\theta$ from folding directions shows excellent agreement with the observed moiré superstructures, in which the $\theta$ can also be directly determined by measuring the periodicity $d$ of the superstructure [$\theta = 2\arcsin(a/2d)$, where $a$ is the graphene lattice constant]. This cross-referencing of the value of $\theta$ provides a check for the STM origami operations and confirms the tunability of twisted stacked GNSs by simply varying the folding direction. Figure 2A (top-right), which summarizes the different experimental values of $\theta$ that have been achieved in the present work, demonstrates the range and the precise control of arbitrary twisting in bilayer graphene made possible by the STM origami.

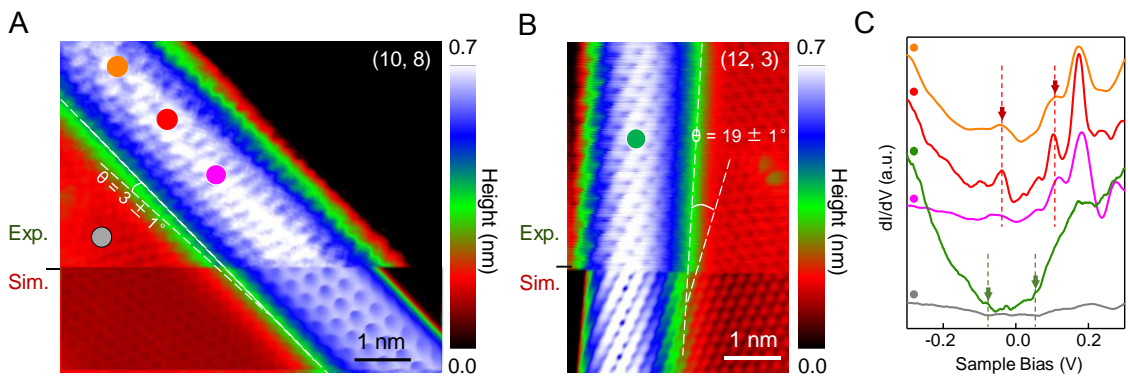

**Fig. 3. Tunable 1D tubular carbon structures with different chirality and electronic properties**. (**A** and **B**) Atomically resolved STM images showing structural configurations of two chiral tubular

structures acquired by folding the same GNI along different directions. The corresponding chiral angle is determined to be 3° and 19° for (A) and (B), respectively. Simulated STM images based on the experimentally-determined structural index (fig. S5) are also provided underneath their corresponding experimental images. (**C**) The *dI/dV* spectra acquired at different locations as labelled in (A) and (B), showing the appearance of VHSs and the distinct electronic properties of two different tubular edges. The color codes of the spectra are kept the same as those of location markers in (A) and (B). The red and green dotted lines are guides to the eye to show the onset of the first VHS peaks (highlighted by arrows, details are shown in fig. S6) of the two folded tubular structures, respectively. For comparison, the data acquired from the flat region (grey) are also presented. Settings: $I_t$ =100 pA, $V_s$ =-0.2 V. T=4.2 K.

In addition to the stacked bilayer nanostructures, folding of a GNI also generically forms a tubular edge whose chirality depends solely on the folding direction (Fig. 1C). It is worth noting that the origami process is essentially the same as the roll-up model of a perfect single-walled carbon nanotube from a monolayer graphene, except that the as-formed tube is not seamlessly closed in the present work (*36*). We have, therefore, employed the conventional chiral-indices notation (*n*, *m*) of a carbon nanotube to define the constructed tubular edges, where *n* and *m* are integers (fig. S5). Figures 3A and 3B show atomic configurations of two tubular edges constructed from the same GNI but using two different folding directions. In Fig. 3A, the angle between the folding axis and hexagonal lattice is about 3 ± 1°, while that in Fig. 3B is 19 ± 1°. Correspondingly, these two single-walled tubes are (10, 8) and (12, 3), respectively (see more detailed analysis in fig. S5). The simulated STM image based on the chiral index assignment is also provided and placed underneath its corresponding STM images for comparison. It should be noted that our simulations of STM images do not consider the tip-convolution effect that often leads to underestimation of tube diameters (*37*). On the other hand, the good agreement between simulated and experimental STM images further confirms our index assignment.

We have also measured *dI/dV* spectra along as-formed tubular edges, and presented data in Fig. 3C. In contrast with the data acquired from the flat bilayer graphene region, a clear manifestation of van Hove singularity (VHS) peaks is observed on both tubular edges, suggesting that, while these tubular edges are not seamlessly closed, they still possess 1D electronic characteristics. Our experimental observation of 1D VHS characteristic from the folded tubes is also corroborated by the density-functional theory (DFT) calculations of electronic structures of both folded tubes and conventional single-wall CNTs (fig. S7) (*38*). Importantly, the consistency of the *dI/dV* spectra acquired along the same tube suggests the delocalized nature of electronic states intrinsic in a 1D structure as well as the defect-free quality of the tubes created by the STM origami. Although the two tubular edges presented in Figs. 3A and 3B are created from the same GNI, they show intriguingly different electronic properties. For example, by comparing their spectra, we find that there exists a small energy shift (31 meV) of the VHS gap from the Fermi energy for the tubular edge in Fig. 3A, which can be attributed to interactions between the tubular edge and the substrate because the (10, 8) tube should behave more as a semiconductor while the (12, 3) tube is more metallic (*39, 40*). These observations highlight the opportunity to investigate effects of the local environment on low-dimensional electronic properties by using the STM origami.

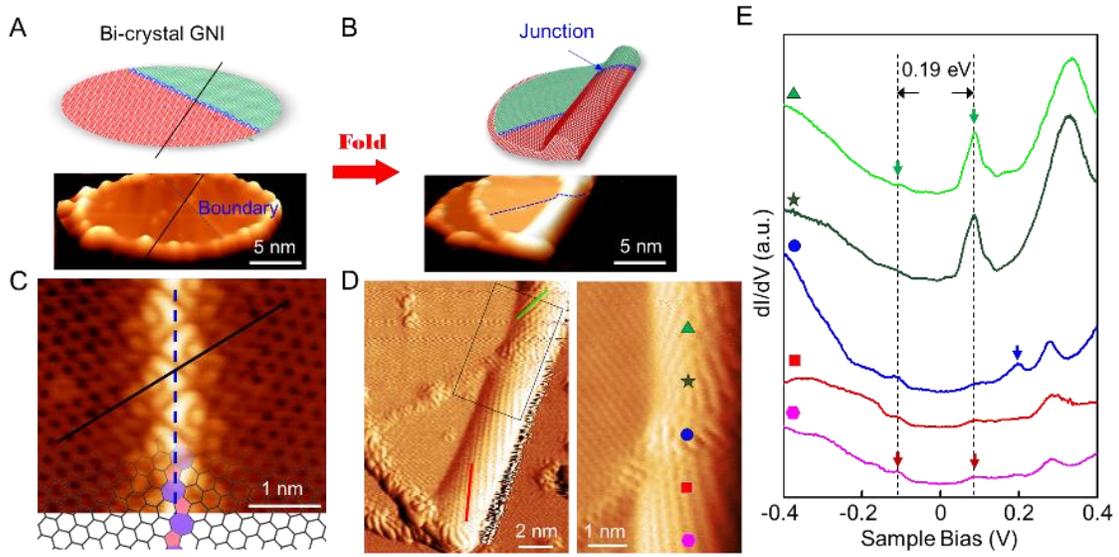

**Fig. 4. Creation of 1D carbon IMJs.** (**A** and **B**) Schematic diagram (top) and STM image in three-dimensional view (bottom) before (A) and after (B) folding a bi-crystal GNI. Red and green shaded regimes represent two different crystalline domains in a bi-crystal GNI with domain boundary (blue dashed line). The black solid line represents the folding axis in the STM origami. (**C**) Atomically-resolved STM image revealing the existence of well-defined 5−7 pairs in the boundary of bi-crystal GNI in (A). A 2D structural model of a GNI with 5−7 pairs (orange pentagons and purple heptagons) boundary is overlaid on the STM image for structural assignment. (**D**) Atomically-resolved STM characterization of the IMJ formed by two folded tubular segments with different chirality (red and green line segments highlight the chiral vectors of the bottom and top tubes, respectively). Left: large-scale; Right: zoom-in image of black rectangular area on the left. (**E**) The *dI/dV* spectra recorded at different locations along the IMJ, labeled by the color symbols in (D). The two doted lines are guides to the eye to show the evolution of the first VHS peaks along the junction. The blue arrow highlights the appearance of a defect state in the junction interface, which is clearly absent at the location away from the IMJ interface. Settings for (A, B): $I_t$ = 10 pA, $V_s$ = -3.0 V; Settings for (C-E): $I_t$ = 100 pA, $V_s$ = -0.2 V.

As we have already have demonstrated, STM origami is a general technique that is not limited to perfect single-crystal GNIs. This immediately opens up an exciting opportunity to create even more complex 2D and 1D GNSs that might be challenging otherwise. For example, 1D carbon IMJs consisting of two different carbon nanotubes seamlessly joined by 5−7 structural defects forming seamless carbon-based size metal-semiconductor, metal-metal and semiconductor-semiconductor building blocks with robust solid-state behavior (*41*), have been proposed to be perfect molecular-scale electronic devices, such as rectifiers, field-effect transistors, switches, amplifiers, photoelectrical devices, etc (*42*). These IMJs have been observed experimentally (*43, 44*), but growth of such IMJs with desirable structural configurations and properties has been intimidating and lacking. We have demonstrated that 1D carbon IMJs between dissimilar tubular edges can be created in a highly control manner by performing STM origami on bi-crystal GNI (Figs. 4A and 4B). Bi-crystal GNIs consisting of two different in-plane graphene domains joined by the well-known 5−7 structural defects (figs. S8 and S9) have been successfully achieved recently (*38*). Because our STM origami is a highly spatially localized technique, it can allow selective folding of bi-crystal GNI across 5−7 pair domain boundaries (Fig. 4C) along different directions, leading to unique construction of edge IMJs with pre-defined structures and properties.

Figure 4D shows one example of such IMJ achieved by STM origami. After folding across a planar 5−7 domain boundary as shown in Fig. 4C, two tubular structures can be clearly resolved with tube indices (9, 4) and (10, 3) for the top and bottom segments, respectively. The difference of chiral vectors between the two segments is about 32º. Accompanying the formation of tubular edges, the 5−7 boundary in the original planar bi-crystal GNI is also folded concordantly, forming a well-defined 1D IMJ interface joining the two tubular segments seamlessly. A series of *dI/dV* spectra are also acquired along the IMJ and presented in Fig. 4E. It can be seen that the energy positions of the first two VHSs and corresponding energy gap (0.19 eV) are almost the same for the two connected (9,4) and (10,3) tubular edges. This feature can be understood by the fact that both tubes possess the same width and noting that the VHS gap is mainly determined by the tube diameter for semiconducting carbon nanotubes (*40*). However, in the junction interface, not only a lattice distortion but also a localized (*dI/dV*) peak at 0.20 eV is clearly observed, which can be attributed to defect states of the 5−7 pairs (*44*). We note that in Fig. 4E, there is a large asymmetry between the two VHS on one side of the junction but not on the other. Such asymmetries are known to exist in CNTs and have been attributed to a variety of effects (Ref. *45* and references therein). Here, differences in lattice deformations are the likely cause.

The emerging IMJ-like structures enabled by STM origami offer a new set of building blocks for demonstrating new physical effects and device concepts. As compared with prior work where IMJs could only be accidentally observed (*42-44*), the present method allows creation of IMJ-like structures from well-defined 5−7 boundaries and different combinations of tubular edges can be integrated in a highly selective manner by simply varying the folding direction. Therefore, the present work provides a route to fabricate novel, complex, and atomically precise carbon nanostructures with engineered electronic properties that may ultimately lead to the construction of graphene-based quantum machines.

**Supplementary Materials:**

Materials and Methods

Supplementary text

Figs. S1 to S9

References (*46-57*)